\documentclass[12pt]{iopart}

\usepackage{graphicx}
\usepackage{svg}
\usepackage{cite}
\usepackage{color}
\usepackage[colorlinks,
linkcolor=blue,       
anchorcolor=blue,  
citecolor=blue,
urlcolor=blue
]{hyperref}
\begin{document}

\title[]{Two-dimensional spatially resolved measurements of helium metastable densities by tunable diode laser absorption spectroscopy in atmospheric pressure RF plasma jets}

\author{David A Schulenberg$^{1}$\footnote[2]{Author is deceased. This work is based on the substantial contributions of our colleague and friend David, who played a leading role in its conception and development. After his unexpected passing, we completed the work and the manuscript in his honor and memory.}, Xiao-Kun Wang$^{1*}$, Máté Vass$^{1,2}$, Ihor Korolov$^1$, Thomas Mussenbrock$^{1}$, Julian Schulze$^{1}$}

\address{$^{1}$ Chair of Applied Electrodynamics and Plasma Technology, Ruhr University Bochum, 44780 Bochum, Germany}
\address{$^{2}$ Institute for Solid State Physics and Optics, HUN-REN Wigner Research Centre for Physics, 1121 Budapest, Hungary}

\ead{xwang@aept.rub.de}
\vspace{10pt}

\begin{abstract}

Helium metastable species play a critical role in sustaining radio-frequency (RF) driven micro atmospheric pressure plasma jets through Penning ionization and for the generation of reactive oxygen and nitrogen species (RONS). Their densities are typically measured using tunable diode laser absorption spectroscopy (TDLAS). Most spatially resolved TDLAS approaches rely on mechanical scanning of a narrow laser beam across the plasma, which is time-consuming and limits spatial resolution. In this work, we present an advanced two-dimensional (2D) TDLAS method that enables direct spatial mapping of helium metastable densities without the need for mechanical scanning. A rotating optical diffuser is employed to suppress speckle interference and generate uniform illumination across the plasma region. The absorption profile is captured using a short-wavelength infrared camera equipped with a telecentric lens, achieving high spatial resolution (approximately 10 $\mu$m) across the entire field of view. This approach significantly enhances both data quality and acquisition speed. The improved 2D TDLAS system is applied to measure helium metastable densities in plasma jets with structured electrodes driven by different tailored voltage waveforms. The results show very good qualitative agreement with fluid simulations and previously reported experimental data.

\end{abstract}


\section{Introduction}
Atmospheric pressure radio-frequency (RF) plasma jets have garnered considerable interest due to their extensive applicability in various fields, including plasma medicine, material surface modification, environmental remediation, and synthesis of nanoparticles \cite{Aboubakr2016,Gibson2014,Adamovich2017,Penkov2015,Reuter2018,Whitehead2019}. Such plasmas are mainly operated in inert gases, with a small admixture of reactive molecular gases, such as nitrogen, oxygen or their mixtures. The widespread use of these plasma jets primarily stems from their ability to generate abundant reactive species, such as reactive oxygen and nitrogen species (RONS) and excited metastable species, under atmospheric pressure and relatively low gas temperature conditions. The unique environment of these jets enables selective chemical reactions that are beneficial for numerous applications involving the processing of temperature-sensitive materials. Among these reactive species, helium metastables are of particular importance as they significantly contribute to the plasma chemistry via processes such as Penning ionization \cite{Niermann_2011, Spiekermeier2014, Niemi2011, Korolov_2020}. The efficient generation and control of helium metastable species directly impact plasma stability, efficiency, and the overall effectiveness of the plasma-based processes. 

Tailored Voltage Waveforms (TVWs) synthesized from low frequency consecutive harmonics were initially proposed in low pressure discharges for the control of ion energy distribution functions (IEDF) at the substrate \cite{Wang2000, Patterson2007}. A much narrower IEDF is achieved compared to the conventional sinusoidal high frequency waveform excitation. Using consecutive high frequency harmonics was found to enable the independent control of the mean ion energy and flux at the substrate via the Electrical Asymmetry Effect \cite{Heil_2008, Schulze_2011}. Recently, the concept of TVWs has also been extended to atmospheric pressure plasma jets, where it has demonstrated promising capabilities in controlling the generation and transport of reactive species by using driving voltage waveforms synthesized from consecutive high frequency harmonics of typically 13.56 MHz to control the space and time dependent Electron Energy Distribution Function (EEDF) \cite{Korolov2021,Korolov2021a,Liu2021b,Vass_2021,Gibson_2019}. A series of experimental and numerical investigations have been conducted to explore these effects in different gas mixtures and configurations. In He/N$_2$, TVWs significantly enhance the energy efficiency of helium metastable production compared to single-frequency operation \cite{Korolov2021}. In He/O$_2$, TVWs enables spatially controlled generation of atomic oxygen, with densities enhanced by increasing the number of harmonics \cite{Korolov2021a}. Hybrid simulations and experiments also confirmed that TVWs induce asymmetric electron heating and enhances the local generation of reactive oxygen species in He/O$_2$ jets \cite{Liu2021b}. Recent computational studies revealed that changing the shape of TVWs allows controlling the EEDF so that, depending on the waveform shape, different neutral species with different energy threshold of their generation by electron-neutral collisions, can be generated more energy efficiently at atmospheric pressure \cite{Vass_2024}. Similar effects of high frequency TVWs were found in low pressure capacitively coupled plasmas, where adjusting the driving voltage waveform shape provides ultimate control of the space and nanosecond time dependent electron power absorption dynamics \cite{Wang_2021,Hartmann_2021} and, thus, improves the energy efficiency and control of radical generation \cite{Wang_2024}. 

In addition to TVWs, structured electrodes have also been used to control electron dynamics \cite{Durian_2022, Schmidt_2013}. This approach was also initially developed in low pressure plasmas, where sheath driven acceleration and confinement of electrons in surface trenches, known as the RF hollow cathode effect, led to localized enhancement of excitation and ionization \cite{Lee2010,Lee2011,Lafleur2012c,Zhang_2025}. Besides enhancing the plasma density, customized structure designs and distributions across large electrode surfaces were found to improve plasma uniformity across large substrates \cite{ohtsu,Wang_STR_electrode}. Although the underlying mechanisms of electron heating differ between low and atmospheric pressure discharges, studies have demonstrated that this method also exhibits clear advantages at atmospheric pressure, where structured electrodes effectively induce local Ohmic fields and current focusing, resulting in enhanced production of helium metastables and atomic oxygen \cite{Liu2023a}.

As the discharge structure becomes more complex, the requirements for plasma modeling also increase accordingly. For systems employing structured electrodes, at least two-dimensional simulations are necessary to resolve the spatial features, which places higher demands on both numerical methods and computational resources. Moreover, experimental diagnostics are essential to validate the simulation results. However, due to the compact geometry of atmospheric pressure plasma jets, only optical and laser-based diagnostic techniques, such as phase resolved optical emission spectroscopy (PROES) \cite{Bischoff2018}, tunable diode laser absorption spectroscopy (TDLAS) \cite{Korolov_2020,niermann2010space}, and two photon absorption laser induced fluorescence (TALIF) \cite{Knake2008} are feasible for measuring the relevant plasma parameters.

For the measurement of Helium metastable densities, TDLAS is commonly employed due to its capability of providing absolute, species-selective measurements. In the works of \cite{Korolov_2020,Spiekermeier2014,Niermann2011}, TDLAS was applied to determine the absorption around 1083 nm, using a diode laser with sub-10 MHz linewidth. The laser beam was reduced and scanned across the plasma jet based on a motorized translation stage, enabling spatially resolved measurements with micrometer precision. The absorption spectra detected by a photodiode were fitted with Lorentzian profiles, and the metastable density was retrieved via the Beer-Lambert law based on the integrated line intensity. However, the conventional TDLAS approach requires point-by-point scanning across the discharge, making spatial mapping time-consuming, especially for two-dimensional (2D) scanning of plasma jets with structured electrodes. This highlights the need for more efficient diagnostic strategies for obtaining high-resolution 2D profiles of metastable species, especially in the case of structured electrode configurations, where strong local gradients in metastable density are expected and accurate spatial resolution becomes particularly critical.

Achieving 1D spatial resolution perpendicular to the electrodes based on recording the laser beam intensity with a camera for the measurements of argon metastable densities using TDLAS has already been demonstrated in low-pressure discharges with a comparatively large electrode spacing of 3 cm \cite{Donko2023}. However, it has not been applied to atmospheric pressure jets with narrow electrode gaps for either one- or two-dimensional spatially resolved measurement yet. In this work, we present a modified approach that enables 2D high-resolution mapping of helium metastable densities in a micro-scale atmospheric pressure plasma jets, even in the presence of structured electrodes. This diagnostic method combines precise optical alignment with a miniaturized scanning scheme, adapted to the geometrical constraints of jet plasmas, and provides detailed insights into the spatial distribution of metastables under various discharge conditions. The experimental results are compared to those of computational simulations.

This paper is structured in the following way: in section \ref{section2}, the experimental setup and diagnostic methods are introduced and the simulation method used in the numerical studies is briefly described. The results are presented in section \ref{section3}. Conclusions are drawn in section \ref{section4}.

\section{Experimental and computational methods}\label{section2}
\subsection{Experimental setup}
\begin{figure}[ht]
	\centering
	\includegraphics[width=0.9\linewidth]{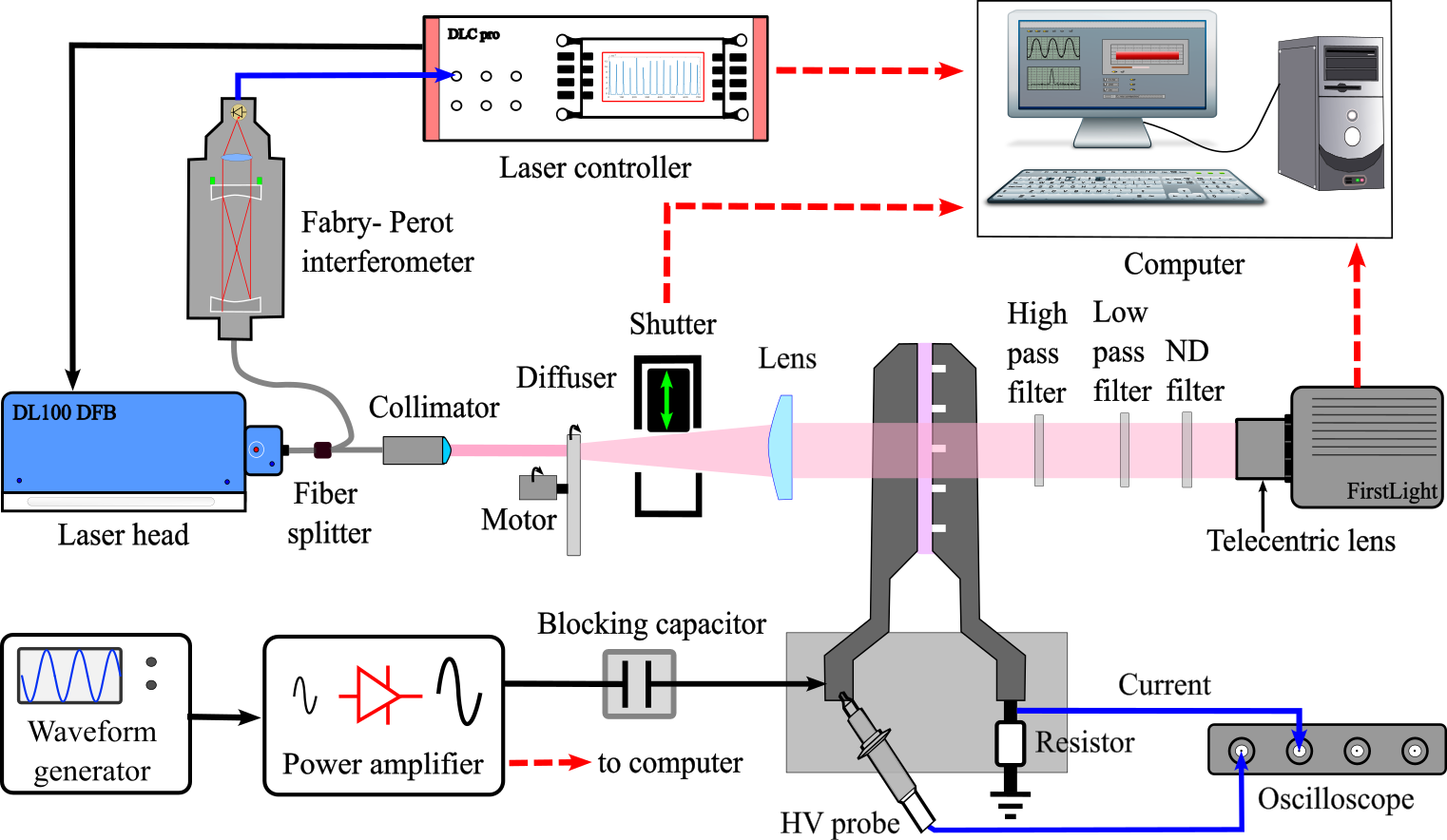}
	\caption{Sketch of the experimental setup with diagnostics.}
	\label{Setup}
\end{figure}

A schematic of the experimental setup along with the TDLAS system is displayed in figure \ref{Setup}. The atmospheric pressure plasma jet used in this work consists of two parallel electrodes with a gap of 1 mm and a discharge region length of 30 mm. The plasma is ignited and sustained in the gap between the electrodes in a gas mixture of helium (5.0 purity) and nitrogen (5.0 purity) with gas flow rates of 1 slm and 0.5 sccm, respectively, which are regulated by mass flow controllers (BronkhorstF-201CV). One of the electrodes has a planar surface and serves as the driven electrode, while the other is a structured electrode featuring five uniformly distributed trenches, each with a depth of 1 mm and a width of 0.5 mm. The powered electrode is driven by sinusoidal or tailored voltage waveforms which are generated using an arbitrary waveform generator (Keysight 33600A) controlled by a LabVIEW-based program on a computer. These waveforms are then amplified using a broadband power amplifier (AR 500A100A) and applied to the driven electrode via a blocking capacitor. As power reflections lie outside the scope of this fundamental study, no impedance matching is implemented for TVWs, and consequently high reflected power is present. The opposite structured electrode is connected to ground. The applied voltage is monitored using a high-voltage probe (Tektronix P6015A with a bandwidth of 75 MHz), while the discharge current is determined by measuring the voltage drop across a 4.7 $\Omega$ resistor connected in series on the grounded side. Both the voltage and current waveforms are recorded by a USB oscilloscope (Picoscope 6402C), which, along with the waveform generator, is fully controlled by the LabVIEW program. To ensure accurate delivery of tailored voltage waveforms, a closed-loop feedback system is implemented. The voltage measured at the powered electrode is first subjected to a Fourier transform to extract the amplitude and phase of each frequency component, which are then directly compared with the corresponding amplitude and phase values of the ideal tailored waveforms. Based on this comparison, the waveform generator output is iteratively adjusted until the measured voltage waveform matches the desired target waveform.

A tunable diode laser source (Toptica, DFB pro L, LD-1083-0070-DFB-1, with digital controller DCL pro) is employed for metastable species diagnostics via absorption spectroscopy. The system is designed to measure the absorption profile of the He-${\rm I}~(2^{3}\mathrm{S}_{1} \rightarrow 2^{3}\mathrm{P}_{J})$ triplet transitions ($J$ = 0,1,2), corresponding to central wavelengths of 1082.909 nm for $J$ = 0, 1083.025 nm for $J$ = 1 and 1083.034 nm for $J$ = 2. The line width of the laser ($ < $2 MHz) is much smaller than the width of the absorption line, which is several GHz due to the strong collisional broadening at atmospheric pressure \cite{Sadeghi2024}. As shown in figure \ref{Setup}, the laser head is equipped with a fiber coupler to guide the laser beam via single-mode optical fibers, and the laser beam is then split by a beam splitter (Thorlabs, TW1064R2A1B) into two parts. 90$\%$ of the beam is directed to a Fabry–Perot interferometer (Toptica FPI 100-0980-3V0 with 1 GHz free spectral range and resolution of 2 MHz) to monitor any unexpected mode hoping and determine the relative laser frequency change. The cavity resonance signal is recorded by an oscilloscope embedded in the laser controller (Toptica DLC pro), which is controlled via a LabVIEW-based control program. The remaining 10$\%$ of the beam passes through a collimator and enters a rotating diffuser (Thorlabs, DG20-220) mounted on a motor (Krick, Max Power 400) driven at 15 V, rotating at approximately 220 Hz. The rotating diffuser is used in the laser beam path to produce a more homogeneous intensity profile and to reduce spatial coherence, thereby preventing interference patterns on optical components. After passing through the rotating diffuser, the diverging laser beam is collimated by a plano-convex lens to produce a uniform, parallel beam that traverses the plasma jet discharge region. Note that the laser power at the collimator exit is approximately 1.18 mW, and at the position of the measured discharge region this corresponds to a laser energy density of about 2.3 W/m$^2$. To check for possible saturation effects, the laser power was varied between 0.59 mW and 1.18 mW. No significant differences were observed in the measurements, indicating that saturation effects can be excluded. The transmitted laser beam then passes through a set of optical filters, including a neutral density filter (OD = 1.0, attenuation factor of 10), a high-pass filter (1050 nm), and a low-pass filter (1100 nm), to isolate the absorption signal, suppress background light and plasma emission, and reduce the laser intensity to avoid saturation of the camera sensor. The signal is finally captured by a short-wavelength infrared (SWIR) camera (First Light, C-Red 2 Lite) equipped with a telecentric lens (Edmund Optics, 63748), enabling spatially resolved absorption measurements. The acquisition speed of the camera is set at 10 PFS and the integration time is 0.006 s. Due to the limited field of view of the lens and the need for high resolution, each image acquisition covers only a single trench. All the results presented in the results were taken from the trench at the center of the electrode.

\begin{figure}[ht!]
    \centering
    \includegraphics[width=0.7\linewidth]{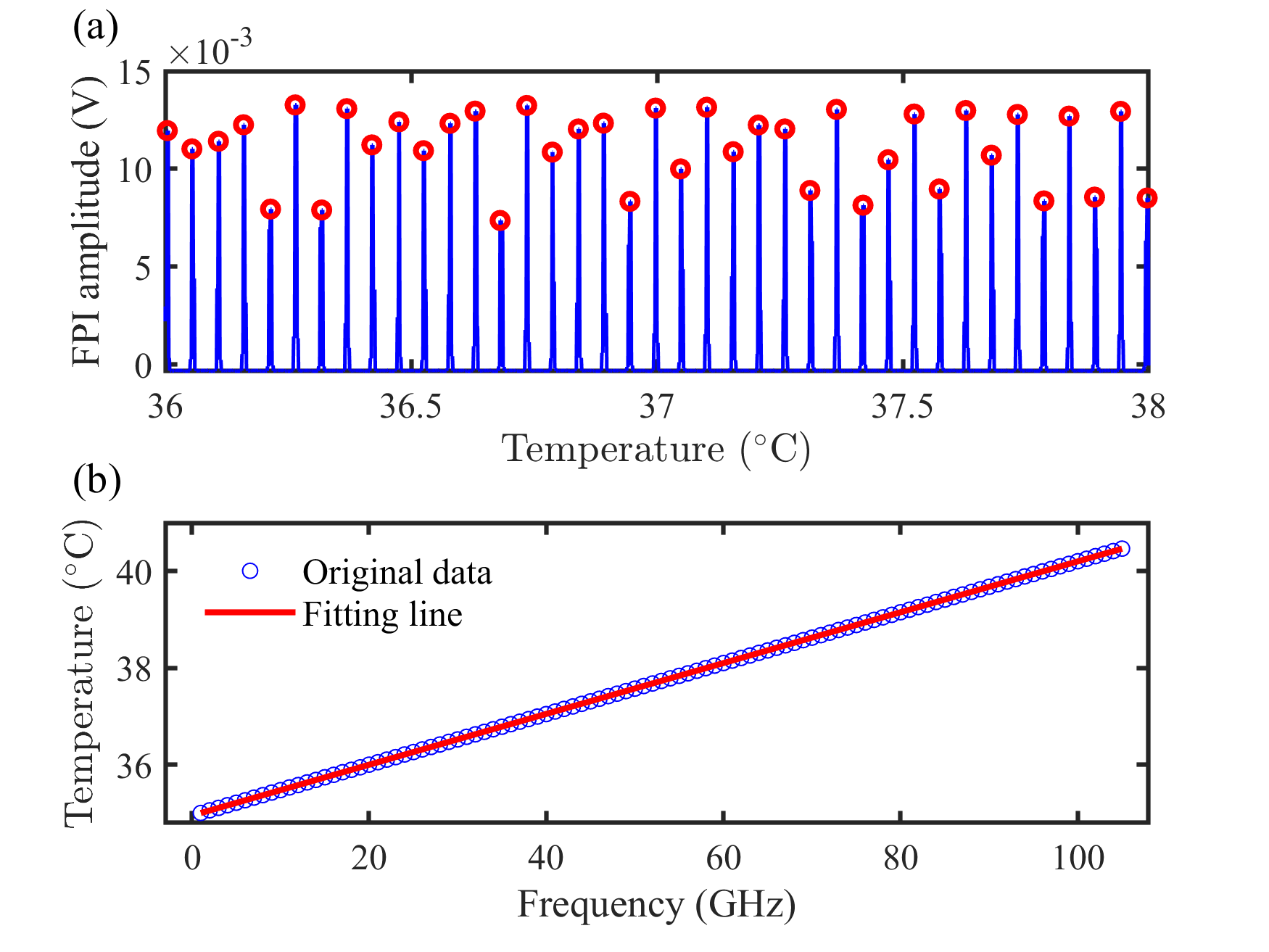}
    \caption{(a) FPI transmission signal with marked cavity resonance as a function of laser diode temperature, and (b) laser diode temperature as a function of laser frequency determined from the FPI signal, with the linear fit shown as a red solid line .}
    \label{FPIdata}
\end{figure}

To obtain the absorption spectral profile, the laser wavelength is tuned by adjusting the operating temperature of the tunable diode laser source. Before performing the absorption measurements, a high-resolution calibration was carried out to determine the relationship between the laser diode temperature and its corresponding optical frequency. The laser temperature was scanned with a fine step size of 0.001 $^{\circ}$C, and the resulting signal recorded by FPI as a function of temperature is shown in figure \ref{FPIdata} (a). It shows a representative portion of the FPI signal within the temperature range of 36$^{\circ}$C to 38$^{\circ}$C, where clear cavity resonance peaks can be observed.  The FPI has a free spectral range of 1 GHz; thus, by identifying the positions of these peaks, the relationship between laser diode temperature and optical frequency can be determined. The full-range relationship between temperature and frequency is plotted in figure \ref{FPIdata} (b), where a linear fit demonstrates an excellent correlation with the experimental data, confirming the reliability and predictability of temperature-based wavelength tuning for the diode laser. During the absorption measurements of the plasma, the temperature was scanned with a larger step size of 0.1$^{\circ}$C, sufficient to provide appropriate sampling across the absorption line profile.

\subsection{Data processing and evaluation procedure}

Based on the Beer-Lambert law, the absolute metastable densities can be obtained by the measured transmittance, $T_\nu$:
\begin{equation}
     T_{\nu}=\frac{I(\nu)}{I_{0}(\nu)}=\frac{I_{\rm Pon}^{\rm Lon}({\nu})-I_{\rm Pon}^{\rm Loff}({\nu})}{I_{\rm Poff}^{\rm Lon}({\nu})-I_{\rm Poff}^{\rm Loff}({\nu})} =e^{-k({\nu})l},
     \label{equ1}
\end{equation}
where $I({\nu})$ and $I_0({\nu})$ are the intensities of the transmitted laser radiation as a function of frequency (${\nu}$) with and without the presence of absorbing species generated in the plasma, respectively. $l$ is the absorption length, which in our setup equals the electrode width of 1 mm. In order to remove the background noise and light, irradiated from the plasma, four signals are recorded, namely the intensity of the transmitted radiation with (1) plasma and laser switched on, $I_{\rm Pon}^{\rm Lon}$, (2) plasma on and laser off, $I_{\rm Pon}^{\rm Loff}$, (3) plasma off and laser on, $I_{\rm Poff}^{\rm Lon}$, and (4) plasma off and laser off, $I_{\rm Poff}^{\rm Loff}$. To reduce statistical noise and improve measurement accuracy, each of the four signals was recorded 5 times at every laser frequency step, and the final values were obtained by averaging the repeated measurements. The entire data acquisition process was automated through a LabVIEW-based control program, which managed the laser shutter and the power amplifier, ensuring synchronized, repeatable, and time-efficient operation. The population density of triplet helium metastables in this work, $n_{2^3{\rm S}_1}$, can be derived from the absorption coefficient, $k({\nu})$ \cite{2004}:
\begin{equation}
    k({\nu})=\frac{e^2}{4\varepsilon _0cm_e}f_{J}n_{2^3{\rm S}_1}F({\nu}),
    \label{equ2}
\end{equation}
where $e$ is the elementary charge, $\varepsilon_0$ is the vacuum permittivity, $c$ is the speed of light in vacuum, $m_e$ is the electron mass and $f_J$ is the absorption oscillator strength. The helium triplet metastable state ($2^3\mathrm{S}_1$) exhibits three closely spaced absorption transitions around 1083 nm, corresponding to the fine-structure components of the $2^3\mathrm{S}_1 \rightarrow 2^3\mathrm{P}_J$ triplet, where $J = 0$, $1$, and $2$. The corresponding oscillator strengths are $f_{J=0}=0.06$, $f_{J=1}=0.18$ and $f_{J=2}=0.3$. $F({\nu})$ is a normalized function ($\int_{0}^{+\infty }F({\nu})d{\nu}=1$) that represents the absorption line shape. Based on equations (\ref{equ1}) and (\ref{equ2}), the absolute line-averaged density, $n_{2^3{\rm S}_1}$, can be determined from the integrated area, $S_J$, under the line-absorption curve, which is given by:
\begin{equation}
     \int_{0}^{+\infty}\ln(T_\nu^{-1})d\nu =S_{J}=\frac{e^2f_Jl}{4\varepsilon_0cm_e}n_{2^3{\rm S}_1}.
    \label{equ3}
\end{equation}

Generally, the shape of the absorption line can be characterized by a Voigt profile, resulting from the convolution of Lorentzian and Gaussian components \cite{corney1978atomic}. The Gaussian width corresponds to Doppler broadening, which arises from the thermal motion of helium atoms and is given by \cite{demtroder1996time}:
\begin{equation}
    \bigtriangleup \nu _{\rm D}=\lambda _{J}^{-1}\sqrt{8\ln2\ k_{\rm B}T_{J}/M_{\rm He}}.
\end{equation}
The gas temperature is within the range of 300 - 350 K inside the jets \cite{Kelly_2015}, the Doppler full width at half maximum (FWHM) is calculated to be approximately 1.72-1.85 GHz. The Lorentzian component, on the other hand, is primarily attributed to pressure broadening at atmospheric pressure, with a typical width on the order of 10 GHz \cite{Niemi_2011}. And other broadening mechanisms, such as natural and Stark broadening, are negligible under the present experimental conditions \cite{Belostotskiy,Yanguas}. Although the absorption line is in principle described by a Voigt profile, the strong collisional broadening causes the three $J$-resolved transitions to merge significantly. Since our analysis relies only on the integrated area of the absorption profile, and fitting tests under comparable conditions have shown that the deviation in integrated area between Voigt and Lorentzian fits is less than 3\%, we use a sum of Lorentzian profiles to approximate the line shape. This simplification enables robust and efficient fitting of the merged triplet structure without introducing significant error.

\begin{figure}[ht!]
    \centering
    \includegraphics[width=0.8\linewidth]{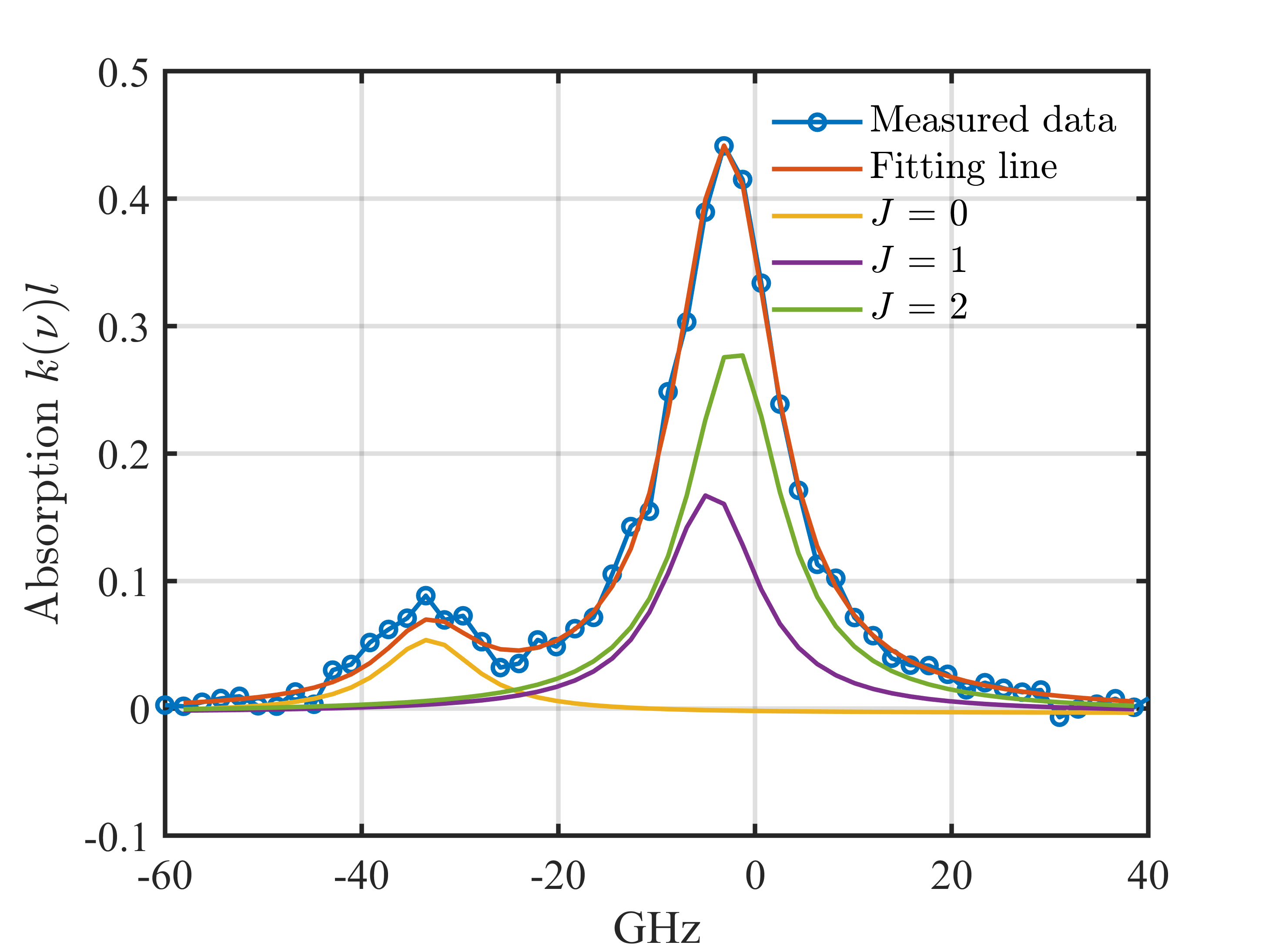}
    \caption{Absorption profile measured by camera within a region of interest in the plasma, shown together with the fitted Lorentzian profiles for the triplet transitions.}
    \label{Absorption}
\end{figure}

Figure \ref{Absorption} shows an example of the absorption line profile measured in a region of interest in the plasma excited by peak-type waveform (as shown in figure \ref{Waveform}) and the fitted results with the sum of three Lorentzian profiles corresponding to the triplet transitions with $J$=0, $J$=1 and $J$=2:
\begin{eqnarray}
f(\nu) &= &f_{b0} 
+ \frac{2 S_{J=0}}{\pi} \cdot \frac{\Delta \nu_{J=0}}{4(\nu - \nu_{J=0})^2 + (\Delta \nu_{J=0})^2} \nonumber\\
&&+ \frac{2 S_{J=1}}{\pi} \cdot \frac{\Delta \nu_{J=1}}{4(\nu - \nu_{J=1})^2 + (\Delta \nu_{J=1})^2} \label{lorentz}\\
&&+ \frac{2 S_{J=2}}{\pi} \cdot \frac{\Delta \nu_{J=2}}{4(\nu - \nu_{J=2})^2 + (\Delta \nu_{J=2})^2},  
\nonumber  
\end{eqnarray}
where $\Delta \nu_J$ and $\nu_J$ are the width and the central position of the corresponding transitions, respectively. $f_{b0}$ is an offset and is typically equal to the experimental noise level. The number of fitting parameters can be significantly reduced by taking into account that $\Delta \nu_{J=0} =\Delta \nu_{J=1} = \Delta \nu_{J=2}$, $\nu_{J=2} = \nu_{J=0} - 31.9$ GHz, $\nu_{J=2} = \nu_{J=1} - 2.3$ GHz, and $S_{J=1}/S_{J=2} = f_{J=1}/f_{J=2}=0.6$, $S_{J=0}/S_{J=2} = f_{J=0}/f_{J=2}=0.2$  \cite{Sadeghi2024}. The Lorentz function can then be expressed as:
\begin{eqnarray}
f(\nu) &= & f_{b0} + \frac{2 S_{J=2} \Delta\nu_{J=2}}{\pi} \nonumber\\  
&& \Bigg( \frac{1}{4(\nu - \nu_{J=2})^2 + (\Delta\nu_{J=2})^2} \nonumber\\
&& + \frac{0.6}{4(\nu - \nu_{J=2} + 2.3\,\mathrm{GHz})^2 + (\Delta\nu_{J=2})^2} \label{fittingline}\\
&& + \frac{0.2}{4(\nu - \nu_{J=2} + 31\,\mathrm{GHz})^2 + (\Delta\nu_{J=2})^2}\Bigg).   
 \nonumber
\end{eqnarray}
It should be noted that a value of 31 GHz is adopted in equation \ref{fittingline} instead of the 31.9 GHz mentioned above. This adjustment yields a significantly better agreement between the fitting line and the experimental data. The observed discrepancy may be attributed to the limited accuracy of the Fabry–Perot resonator, which is specified to be within $\pm$1\%, while in this case, the deviation reaches approximately 3\%. Alternatively, it could also result from a pressure-induced shift of the line, which may differ from those of the other transitions.

Since the full absorption profile is captured during the measurements, integrating over the entire profile yields the total area $S_J$ corresponding to all triplet transitions. Therefore, equation \ref{equ3} can be rewritten as:
\begin{equation}
     S_{J}=S_{J=0}+S_{J=1}+S_{J=2}= \frac{e^2(f_{J=0}+f_{J=1}+f_{J=2})l}{4\varepsilon_0cm_e}n_{2^3{\rm S}_1}.
    \label{totalS}
\end{equation}
In this case, instead of fitting, we can directly obtain the total area $S$, from which the metastable density is determined. Both methods yield consistent results within the statistical error. The choice between fitting and integration depends on the noise level and whether the full profile is available, as this affects the measurement time. In principle, if the profile shape is known and the gas temperature and pressure are assumed to be uniform along the jet, measuring only at the position of maximum absorbance is sufficient to determine the metastable density, allowing for a significant reduction in measurement time.

\subsection{Computational method}

\begin{figure}[ht!]
    \centering
    \includegraphics[width=\linewidth]{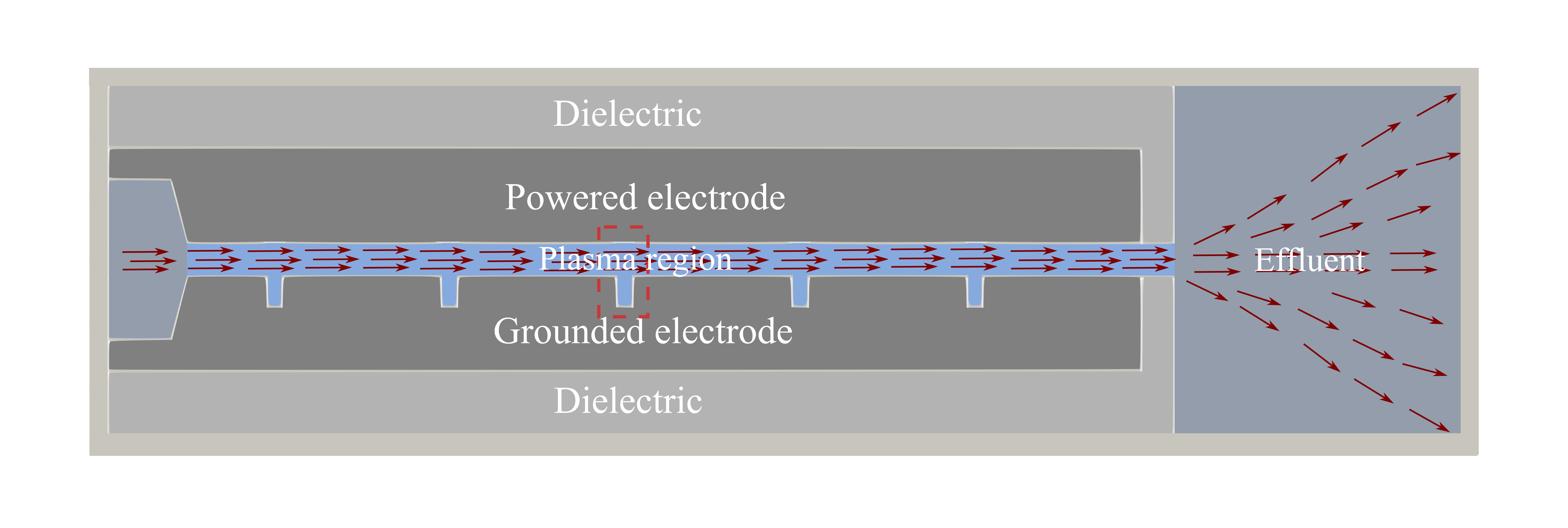}
    \caption{Schematic of the simulated two-dimensional domain. The area shaded in blue indicates the plasma region. The red arrows indicate the direction of the gas flow for illustrative purposes and do not represent simulated flow fields. }
    \label{fig:mesh}
\end{figure}

The two-dimensional plasma fluid simulations were performed using the {\it nonPDPSIM} code \cite{norberg2015formation}. The model self-consistently solves the continuity equations for charged and neutral species using the drift-diffusion approximation, along with Poisson’s equation for the electric potential, on an unstructured triangular mesh. Electron transport and rate coefficients are derived from a two-term Boltzmann solver and applied via the local mean energy approximation, using the solution of the electron energy balance equation. The gas dynamics are described by a compressible Navier–Stokes solver. The computational domain, shown in figure~\ref{fig:mesh}, consists of 16,500 mesh nodes, with a minimum spatial resolution of $5\cdot10^{-5}$~m. Note that in the results section, only the simulation outputs from the region marked by the red dashed rectangle in figure~\ref{fig:mesh} are presented, so as to enable direct comparison with the corresponding experimental results. The dielectric surrounding the electrodes is quartz glass, characterized by a relative permittivity of $\epsilon_{\rm r}=4$. The simulation employs adaptive time stepping, with a mean time step of approximately 0.1 ns.

The charged species in the He/N$_2$ model include electrons (e$^-$) and positive ions (He$^+$, He$_2^+$, N$_2^+$, N$_4^+$). The neutral species considered are He, N$_2$, and the helium metastables He(2$^3$S) and He(2$^1$S). The reaction set (comprising Penning ionization, ion conversion, and recombination) along with the cross sections used in the Boltzmann solver and the surface coefficients for charged species (including electron reflection and secondary electron emission), are identical to those reported in \cite{donko2022vacuum}. For helium metastables, a sticking coefficient of unity is assumed \cite{Liu2021b}.

Due to the low nitrogen concentration used in the experiment (0.05\%), the electron energy relaxation length can become significant at high energies (up to $\sim$0.05~mm), rendering pure fluid simulations less accurate \cite{vass2024new}. To compensate for this discrepancy, simulations typically require a higher voltage amplitude than used experimentally to achieve comparable results \cite{liu2021electron}. In this work, a peak-to-peak voltage of 700~V is applied in the simulation, as opposed to the 500~V used in the experiments.

\section{Results} \label{section3}

\begin{figure}[ht]
	\centering
	\includegraphics[width=0.6\linewidth]{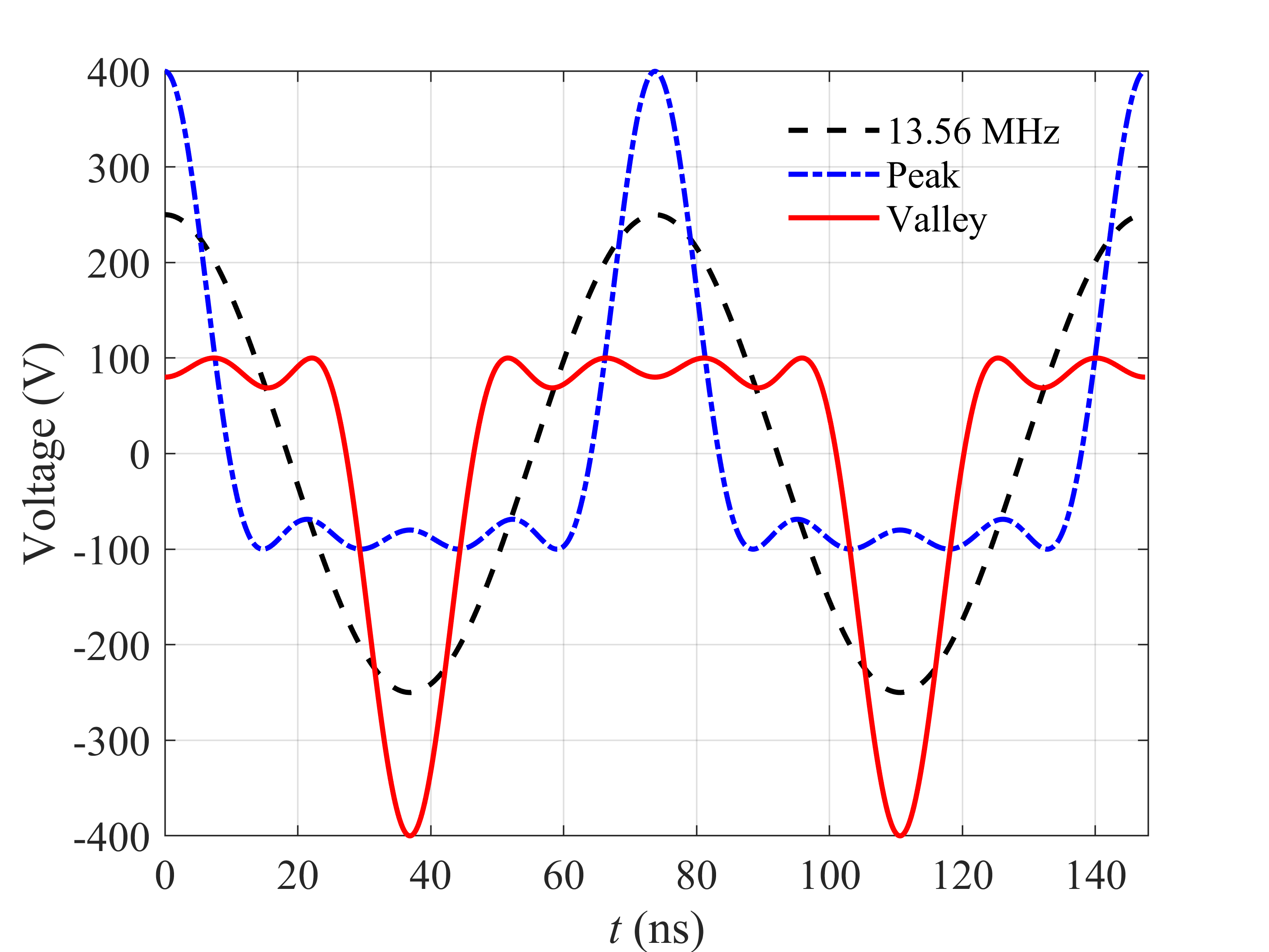}
	\caption{Driving voltage waveforms: sinusoidal waveform with a frequency of 13.56 MHz, Peak- and Valley-type waveforms at constant peak-to-peak voltage value of 500 V.}
	\label{Waveform}
\end{figure}

\begin{figure}[ht!]
    \centering
    \includegraphics[width=1.0\linewidth]{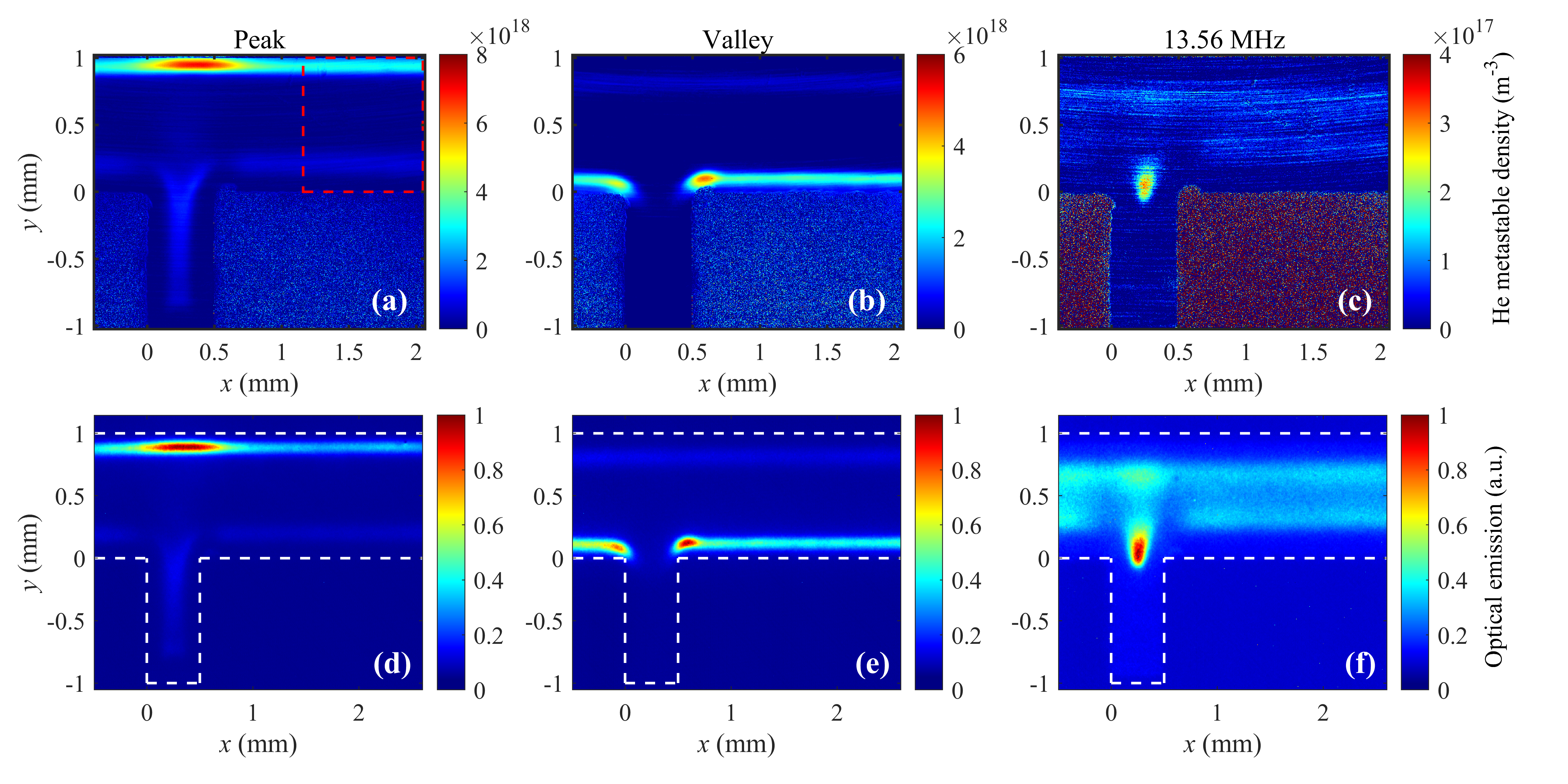}
    \caption{Experimentally measured 2D spatial distributions of the (triplet) Helium metastable density (first row) and of the time averaged optical emission at 706.5 nm (second row) in the jet plasma excited by peak-type (first column), valley-type (second column) and sinusoidal (third column) voltage waveforms with the peak to peak voltage value of 500 V near the middle trench. The gas flow rate of He is 1 slm with an admixture of N$_2$ of 0.5 sccm, directed along the x-axis.}
    \label{Hedensity2D}
\end{figure}

To perform the 2D TDLAS measurements, three types of voltage waveforms are employed, all with a base frequency of 13.56~MHz and a peak-to-peak voltage of 500~V: a single-frequency sinusoidal waveform, and peaks- and valleys-type tailored voltage waveforms, as shown in figure~\ref{Waveform}. The tailored voltage waveforms are synthesized by superimposing the 13.56 MHz fundamental frequency with its first 3 consecutive harmonics. The He metastable densities are measured at the location near the middle trench under the three driving voltage waveforms, as shown in the first row of figure \ref{Hedensity2D}. It shows the two-dimensional spatial distribution of the measured He metastable density, captured with high spatial resolution. Based on the optical setup consisting of the imaging lens with magnification by a factor of 4 and the camera with a 15 $\mu$m pixel pitch, the spatial resolution of the system is approximately 3.75 $\mu$m per pixel. However, the actual resolution is affected by optical elements such as the diffuser, plano-convex lens, and the quartz sidewalls of the jet, all of which introduce additional beam divergence or scattering. Taking these factors into account, the effective resolution is conservatively estimated to be approximately 3 times the pixel resolution, i.e., around 10 $\mu$m. This resolution is sufficiently high for the clear visualization of fine structures, particularly the localized peak regions of the He metastable density.

In contrast, a previous approach reported in \cite{Korolov_2020} relied on moving the plasma jet relative to a fixed laser beam, using a step size of 20 $\mu$m for the spatial resolved measurement of He metastable density, which effectively limits the spatial resolution. Moreover, due to the need to avoid Fabry-Perot resonances in the quartz glass of the jet, the laser beam was introduced at an oblique angle (73$^\circ$) relative to the direction of gas flow, which prevented accurate probing in trench regions. These limitations are inherently avoided in the current imaging-based 2D TDLAS configuration, since the laser beam is not coherent and no Fabry–Perot resonances occur within the optical path. The laser beam is expanded to illuminate the full cross section of plasma and spatial absorption is recorded simultaneously over the entire field of view, enabling accurate two-dimensional mapping even in complex geometries.

As seen in the first row of figure \ref{Hedensity2D}, local enhancements of the He metastable density can be observed for different voltage waveforms near the trench. For the peak-type voltage waveform case in figure \ref{Hedensity2D}(a), the maximum of the He metastable density is above the trench close to the upper electrode, since the local sheath is expanded for most of the fundamental RF period and, thus, Penning ionization and electron impact excitation are strong at this electrode. A maximum of the metastable density at the top electrode is located above the trench immersed into the bottom electrode, since the presence of the trench locally enhances the generation of energetic electrons via the current focusing effect \cite{Liu2023a}. Note that this maximum is not located directly above the trench, but appears further downstream because of the gas flow. In addition, the density distribution penetrates into the trench, with elevated concentrations at the trench opening. This feature is attributed to current focusing effects due to the local field intensification. For the valley-type voltage waveform case in figure \ref{Hedensity2D}(b), the high helium metastable density region shifts to the lower electrode, where the sheath is now expanded for most of the fundamental RF period and, thus, Penning ionization and electron impact excitation are strong at this electrode. In the vicinity of the trench, the distribution is different from the peak case, since it is governed by Penning ionization effects and local field enhancements at the trench orifice. A higher He metastable density is observed on the downstream side (right) of the trench opening. A similar phenomenon has been observed in single frequency excited He/O$_2$ plasma jets with structured electrodes before \cite{Liu2023a}. It is because the gas temperature inside the trench is enhanced, leading to a decrease of the neutral gas densities of He and N$_2$ according to the ideal gas law in the downstream region. Thereby, electrons can gain more energy in this downstream region being accelerated by the RF electric field over a longer mean free path. For the sinusoidal case, the He metastable density is also enhanced, but at the center of the trench orifice as a consequence of the focusing effect of the conduction current density, i.e. electrons inside the trench are pushed together by the expanding sheaths at the two side walls and the bottom boundary of the trench. These electrons are accelerated to high energies at the trench orifice, and then excite ground state helium atoms to the metastable state via collisions. However, the enhancement of the He metastable density under sinusoidal excitation is weaker than that observed in the peak- and valley-type waveform cases, which is consistent with findings reported in previous studies\cite{Korolov_2020}. 

As He metastables are generated through collisions between energetic electrons and ground-state helium atoms, their density profiles are correlated with the distribution of the energetic electrons. As shown in the second row of figure \ref{Hedensity2D}, the spatial distributions of the optical emission at 706.5 nm measured by phase resolved optical emission spectroscopy (PROES) exhibit a high degree of similarity to the distribution of the He metastable density, thereby validating the accuracy and physical interpretation of the metastable density profiles.

\begin{figure}[ht!]
    \centering
    \includegraphics[width=1.0\linewidth]{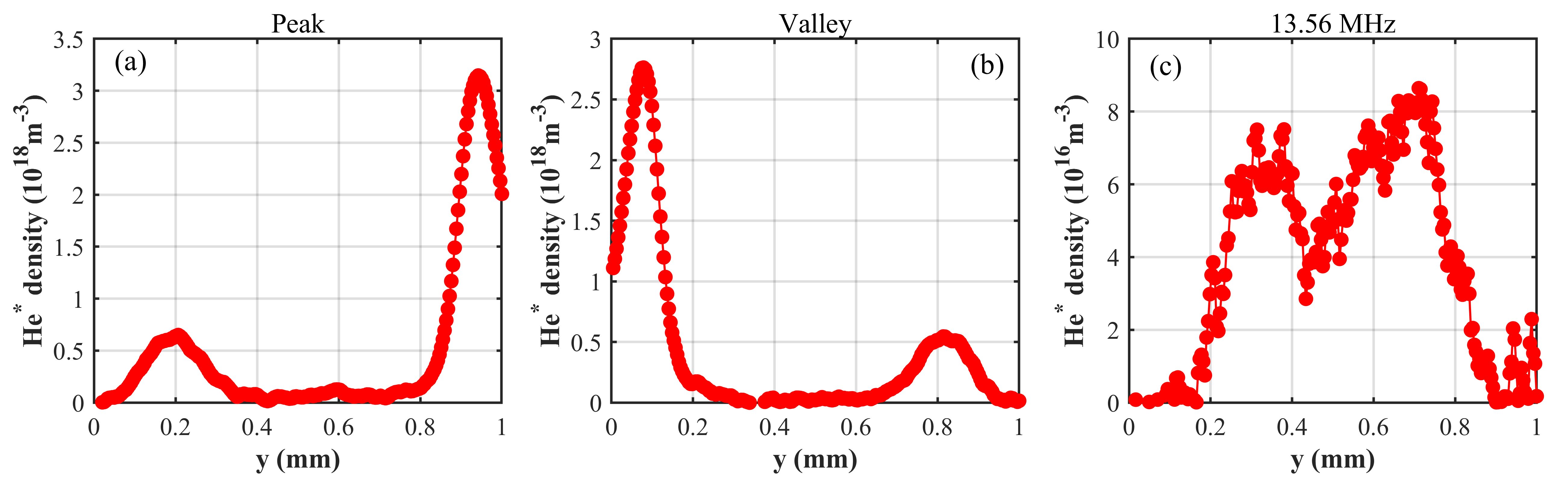}
    \caption{Helium metastable density density profile with spatial resolution perpendicular to the electrodes obtained by spatially averaging the 2D resolved TDLAS measurements in x-direction above the planar electrode segments (which is marked by red dashed rectangle in figure \ref{Hedensity2D}(a)) in plasmas excited by (a) peak-type, (b) valley-type and (c) sinusoidal voltage waveforms with a peak-to-peak voltage of 500 V. The gas flow rate of He is 1 slm with an admixture of N$_2$ of 0.5 sccm.}
    \label{DensityProfile}
\end{figure}

Figure \ref{DensityProfile} presents the He metastable density profiles obtained by spatially averaging the two-dimensional density distribution (shown in figure \ref{Hedensity2D}) along the x-direction in the region to the right of the trench (which is marked by red dashed rectangle in figure \ref{Hedensity2D}(a)), where no significant structural perturbation is present. Such averaging is done to compare the measurements obtained by the new 2D TDLAS diagnostic with those of previously reported one-dimensional measurements \cite{Korolov_2020}. The maximum metastable density in the present study is about 3.2 $\times$ 10$^{18}$ m$^{-3}$ in the presence of a peak-type voltage waveform, which is lower by about 30$\%$ compared to the 4.7 $\times$ 10$^{18}$ m$^{-3}$ reported previously under the same conditions. This discrepancy likely arises from (1) the use of the trenched electrodes (with measurements taken only within the planar electrode regions), (2) differences in electrode surface contamination or degree of usage, and (3) the degree of gas purity and the overall cleanliness of the jet gas handling system. Additionally, in the single frequency case, a small but noticeable left–right asymmetry in the metastable density profile is present. It primarily arises from the discharge asymmetry caused by the asymmetric electrode structure. The curve in (c) appears less smooth than those in (a) and (b) because, in the single-frequency case, the metastable density is more than an order of magnitude lower. As a result, the signal is closer to the detection limit of the current system, which is approximately in the range of $10^{16}$ m$^{-3}$, leading to increased noise in the data. Overall, the results remain within a reasonable margin compared to previous results obtained by 1D TDLAS, but the spatial resolution is significantly improved.

\begin{figure}[ht!]
    \centering
    \includegraphics[width=1.0\linewidth]{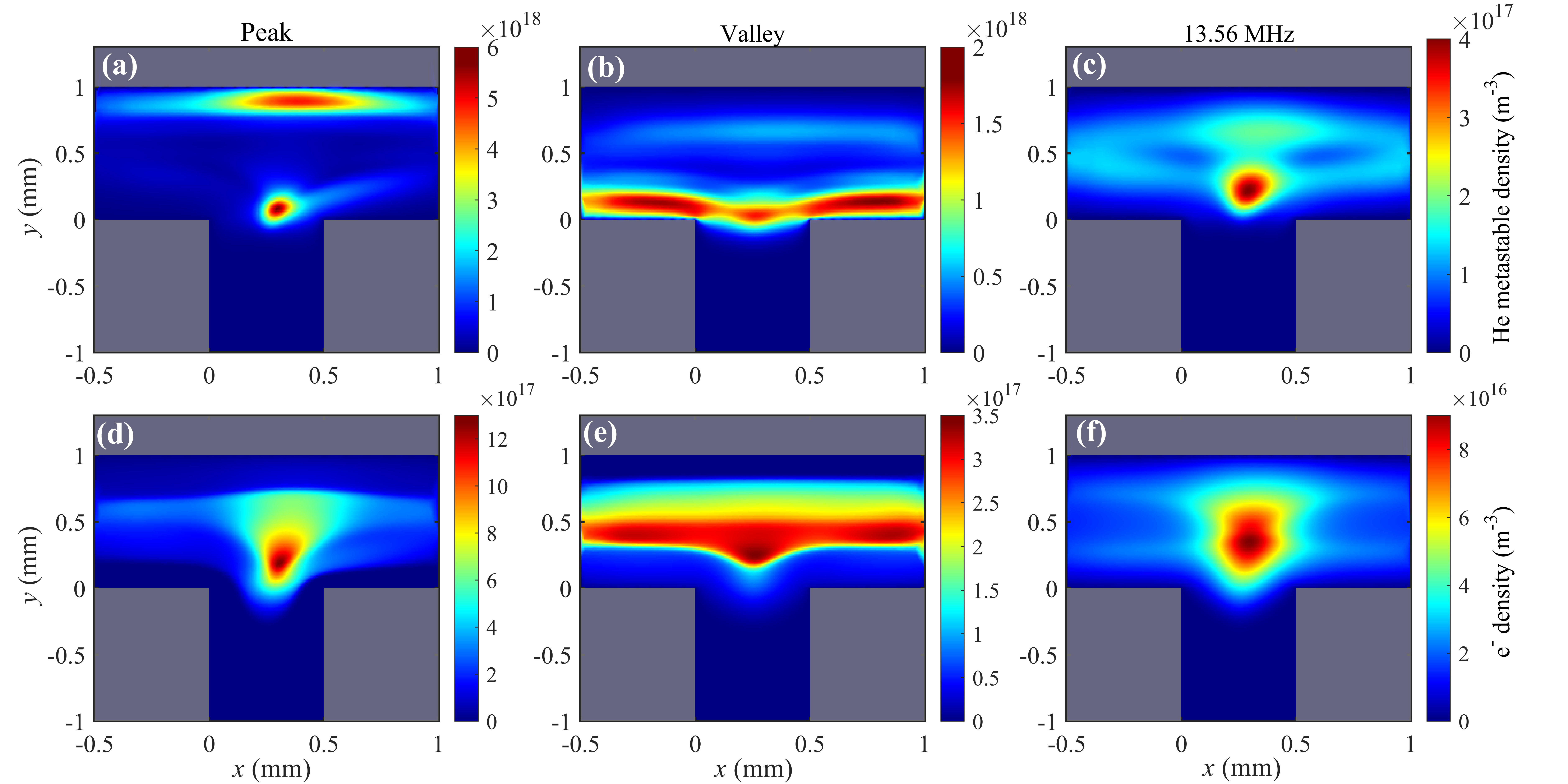}
    \caption{Simulation results for spatial distributions of the (singlet $+$ triplet) Helium metastable density (first row) and electron density (second row) in the jet plasma excited by peak-type (first column), valley-type (second column) and sinusoidal (third column) voltage waveforms with a peak-to-peak voltage value of 500 V. The gas flow rate of He is 1 slm with an admixture of N$_2$ of 0.5 sccm.}
    \label{SimPeak}
\end{figure}

Figure \ref{SimPeak} shows the fluid simulation results of the spatial distribution of the He metastable density and the electron density. Note that the simulated He metastable density includes singlet and triplet states. As the generation rate of helium triplet metastables is approximately three times higher than that of singlets because of the differences of their reaction cross sections and energy thresholds \cite{Korolov_2020}, the triplet state represents the dominant species. The simulated spatial distribution of the He metastable density captures the main characteristics observed in the experimental measurements. Under the peak-type voltage waveform condition, localized enhancements in He metastable density are observed at the trench opening and above the trench opening, near the upper electrode. This feature corresponds well with the electron density distribution shown in the second row, where a higher electron density appears around the trench entrance. The increase in electron density near the trench opening is attributed to enhanced ionization by high-energy electrons accelerated by the sheath electric field near the trench sidewalls. Simultaneously, electron-impact excitation reactions are intensified in this region, leading to the generation of He metastable atoms and thus a local density peak occurs at the trench opening. Additionally, as shown in figure \ref{SimPeak} (d), the electron density above the trench is enhanced and the ion density as well as the ion flux toward the upper electrode are similarly increased. The secondary electrons induced by ion bombardment on the upper electrode are accelerated by the sheath electric field and excite a large number of He metastable states, creating a pronounced He metastable density peak adjacent to the upper electrode. For the valley-type voltage waveform, the He metastable density enhancement region is observed to shift downward, moving closer to the lower electrode. This spatial redistribution is consistent with the dynamics of electron acceleration and excitation dynamics under valley conditions. In the case of the 13.56 MHz sinusoidal excitation, both the He metastable density and electron density are enhanced at the trench opening because of the current focusing effects and the enhanced secondary electron effects caused by the elevated ion flux to the upper electrode, which also agrees well with the experimental results.

Despite the good agreement overall, discrepancies between the simulation and experimental results are observed. In the simulation, neither electrons nor He metastable species significantly penetrate into the trench interior. This is primarily due to the inherent limitations of the fluid model, which is not capable of capturing the localized kinetic behavior of high-energy electrons accelerated by the local electric field induced by the trench geometry. These results suggest that a kinetic simulation approach would be more suitable for accurately resolving the physical processes within the trench. Nevertheless, the fluid model used in this study is sufficient for capturing the major features of the He metastable distribution.

\section{Conclusions}\label{section4}
In this study, we successfully applied tunable diode laser absorption spectroscopy (TDLAS) to achieve high-resolution two-dimensional spatial mapping of the triplet helium metastable atom (He-$\mathrm{I}~2^3{\rm S}_1$) density in an atmospheric pressure COST jet. This diagnostic technique enables direct, non-intrusive, and 2D spatially resolved measurements of He-$\mathrm{I}~2^3{\rm S}_1$ metastable density distributions, offering significant advantages for characterizing the internal structure and dynamics of atmospheric pressure plasma jets.

While the TDLAS technique has previously been employed for 1D metastable He density measurements, it was typically implemented by mechanically scanning the plasma jet with a narrow laser beam and collecting transmitted signals using a photodiode. This method, however, results in long acquisition times and limited spatial resolution due to the laser beam size and angled alignment between the laser and jet in the direction of gas flow to avoid Fabry-Perot resonances in the quartz plates. In this study, we introduced a rotating optical diffuser combined with a camera equipped with a telecentric lens. The diffuser effectively reduces laser coherence and eliminates speckle patterns, resulting in uniform illumination across the plasma region. This approach provides significantly improved spatial resolution determined by the camera-lens parameters and dramatically shortens acquisition times by capturing the entire absorption profile simultaneously.

Using the improved TDLAS system, we measured the He-$\mathrm{I}~2^3{\rm S}_1$ metastable density distributions in a jet with a trench structured electrode in the presence of different driving voltage waveforms: peak-, valley-type and sinusoidal voltage waveforms. Under the peak-type waveform, the He-$\mathrm{I}~2^3{\rm S}_1$ density exhibits a strong localized enhancement above the trench opening and close to the upper planar electrode. For the valley-type voltage waveform, the enhancement of the He-$\mathrm{I}~2^3{\rm S}_1$ density shifts downward with a higher density on the downstream side (right) of the trench opening. Under 13.56 MHz single frequency excitation, the enhancement of the He-$\mathrm{I}~2^3{\rm S}_1$ density appears near the trench opening. The measurement results capture all the details of the metastable density distribution near the trench electrode and are consistent with fluid simulation results.

Overall, this work demonstrates the effectiveness of the 2D TDLAS diagnostic technique for resolving complex spatial structures in atmospheric pressure plasmas jet, making it a powerful tool for the analysis of fundamental physical mechanisms and discharge optimization in microplasma systems.

\section*{Acknowledgments}
 This work was supported by the German Research Foundation in the frame of the collaborative research center SFB 1316, Projects A4 and A5 as well as project MU 2332/12-1. The authors thank Prof. Mark Kushner for providing the {\it nonPDPSIM} code.

\section*{Data availability statement}
The data that support the findings of this study are available
upon reasonable request from the authors.

\section*{References}

%\bibliography{references}
%\bibliographystyle{iopart-num}
\providecommand{\newblock}{}

\end{document}